\documentclass[%
 reprint,
superscriptaddress,
showpacs,
showkeys,
 amsmath,amssymb,
 aps,
 pra,
]{revtex4-2}
\usepackage{graphicx}
\usepackage{epstopdf}
\usepackage{dcolumn}
\usepackage{bm}
\usepackage{multirow}
\usepackage{romannum}
\usepackage{array}
\usepackage[utf8]{inputenc}
 \usepackage{url}
 \usepackage{xcolor}
\usepackage{floatrow}
 \usepackage{caption}
 \usepackage{lipsum}
\usepackage{pgfplots}
\pgfplotsset{compat=1.18} 
\usepackage{pgfplotstable}
\usepackage{xcolor}
\usepackage{float}
\usepackage[margin=1in]{geometry}
\usepackage{subcaption}
\usepackage{adjustbox}
\usepackage{tikz}
\usetikzlibrary{patterns}
 \usepackage{url}
\usetikzlibrary{pgfplots.groupplots}
\DeclareUnicodeCharacter{0327}{ '} \usepackage[justification=raggedright,singlelinecheck=false]{caption}
\usepackage{verbatim}
 \usepackage{url}
\pgfplotsset{compat=1.18}
\usepackage{amsmath}
\usepackage{quantikz}
\usepackage{siunitx}
\usepackage{soul}
\usepackage[dvipsnames]{xcolor}
\usepackage{titlesec}
\usepackage{hyperref}

\titleformat{\subsubsection}
  {\normalfont\normalsize\bfseries\filcenter} 
  {\thesubsubsection.}               
  {1em}                             
  {}                                

\begin{document}

\preprint{APS/123-QED}

\title{Machine Learning-Based Quantum Error Mitigation for Bound-State Simulations}

\author{Sumit Choudhary}
\affiliation{Department of Physics, Guru Nanak Dev University, Amritsar 143005, India}

\author{Tarunpreet Kaur}
\affiliation{
Khalsa College of Engineering and Technology,
Amritsar 143001, India
}
\author{Harjeet Kaur}
\affiliation{Department of Physics, Guru Nanak Dev University, Amritsar 143005, India}

\author{Harpreet Singh}
\email[Corresponding author: ]{harpreet.phy@gndu.ac.in}
\affiliation{Department of Physics, Guru Nanak Dev University, Amritsar 143005, India}

\begin{abstract}
Noise in quantum devices remains a major challenge for reliable quantum simulations, making quantum error mitigation an important tool for improving their accuracy. Machine learning-based quantum error mitigation (ML-QEM) provides a useful approach, but its performance depends on the training datasets and may produce reconstructed quantum states that are not physically valid. In this work, we introduce an ML-QEM framework that uses residual learning with convex projection to ensure that reconstructed density matrices satisfy physically valid density matrices. We study the effect of different training datasets and regression models using Haar-random and variational quantum eigensolver (VQE) ansatz states. We compare the proposed approach with conventional ML-QEM, readout error mitigation, and zero-noise extrapolation. The method is tested on IBM quantum hardware for three- and four-qubit simulations of the bound $\Lambda$-deuteron and $\Lambda$-$\alpha$ hypernuclear systems, respectively. For the three-qubit system, ridge regression trained on VQE-ansatz states gives a ground-state energy of $-0.134$ MeV, close to the ideal value of $-0.135$ MeV, with a state fidelity of $0.999$. For four-qubit system, kernel ridge regression gives ground-state energies of $-3.014$ and $-3.045$ MeV, with corresponding state fidelities of $0.995$ and $0.998$. These results show that physically constrained reconstruction, along with the choice of training datasets and regression models, can improve ML-QEM performance for the quantum simulations studied in this work. 
\end{abstract}

\maketitle


\section{\label{1}Introduction}
Quantum computing has the potential to solve computationally complex problems that are hard for classical computers\cite{jaeger2007classical,preskill2018quantum}. However, current quantum computers are affected by noise arising from decoherence, imperfect quantum gates, and measurement errors, which limit the accuracy of computations performed on noisy intermediate-scale quantum (NISQ) devices \cite{resch2021benchmarking}.

Quantum error correction is the method for reliable computations, but it requires fault-tolerant quantum hardware with number of physical qubits to encode a single logical qubit, makes it difficult to use on current quantum devices \cite{devitt2013quantum,knill1997theory}. Quantum error mitigation (QEM) provides an alternative way to improve computational accuracy without requiring logical qubits or fault-tolerant operations \cite{qin2022overview,cai2023quantum}. It uses measurements from noisy quantum circuits to estimate results that are closer to the ideal, noise-free values.

Several QEM methods have been developed to reduce the effects of noise in quantum computations \cite{li2017efficient,temme2017error,endo2018practical,cai2023quantum,giurgica2020digital}. These methods use measurements from repeated executions of noisy circuits to estimate the noise-free expectation values of observables. Two commonly used methods are zero-noise extrapolation (ZNE) \cite{li2017efficient} and probabilistic error cancellation (PEC) \cite{van2023probabilistic}. However, these methods can require a large number of measurements and, in some cases, information about the noise in the quantum circuit. This increases their computational cost and makes them difficult to use for larger problems \cite{tsubouchi2023universal,takagi2022fundamental,takagi2023universal,quek2024exponentially,schultz2022impact}.

Machine learning-based quantum error mitigation (ML-QEM) has emerged as another approach to reduce errors in NISQ devices \cite{strikis2021learning, liao2024machine, czarnik2021error, czarnik2025improving, bennewitz2022neural, kim2020quantum, muqeet2024machine}. These methods learn the effects of noise from training data and can reduce errors without requiring prior knowledge of the noise model. However, their performance majorly depends on the training dataset and the regression model. Also, the density matrices reconstructed from the predicted Pauli coefficients may not represent valid quantum states. Therefore, two important issues need to be addressed in this conventional ML-QEM: the choice of suitable training states for a given quantum simulation and the physical validity of the reconstructed density matrices.

The choice of training states can significantly affect the performance of ML-QEM. Haar-random states provide a general way to generate a training sample dataset in the Hilbert space. However, they may not represent the states relevant to a specific quantum simulation. Another approach is to generate training states using the same variational circuit ansatz used in the target calculation. In this work, we compare two training strategies based on Haar-random states and VQE-ansatz states generated using the circuit family employed in target simulations.

A reconstructed density matrix must satisfy Hermiticity, unit trace, and positive semidefiniteness to represent a physically valid quantum state. Several methods have been developed in quantum state tomography to reconstruct density matrices that satisfy these conditions \cite{banaszek1999maximum,blume2010hedged,huszar2012adaptive, opatrny1997least, kaznady2009numerical, qi2013quantum}. Maximum likelihood estimation \cite{hradil1997quantum, baumgratz2013scalable, singh2016constructing, schwemmer2015systematic, shang2017superfast} and semidefinite programming-based convex optimization \cite{gaikwad2021true,branderhorst2008optimal, huang2020reconstruction} are commonly used for this purpose. In this work, we use the exact Frobenius-norm projection onto the set of valid density matrices \cite{smolin2012efficient}. The projection is carried out by projecting the eigenvalues onto the probability simplex using a sorting-based procedure \cite{wang2013projection,duchi2008efficient}. We combine this projection with residual learning in ML-QEM to ensure that the reconstructed density matrices satisfy the physical constraints.

We apply the proposed ML-QEM framework to estimate the ground-state energies of three- and four-qubit mapped Hamiltonians describing the bound $\Lambda$-deuteron ($^{3}\mathrm{H}_\Lambda$) and $\Lambda$-$\alpha$ ($^{5}\mathrm{He}_\Lambda$) hypernuclear systems, respectively. We first perform the calculations using the quantum-classical hybrid VQE algorithm on an IBM Quantum processor. We then introduce the proposed ML-QEM framework and examine the effects of the different training dataset, regression model, and physical constraints on the error-mitigation performance. We use two training strategies based on Haar-random states and VQE-ansatz states, with 2000 training states generated for each strategy. Noisy measurements are obtained from the quantum processor, whereas ideal expectation values are calculated using a statevector simulator. The reconstructed density matrices are projected onto the set of physical states to ensure that they satisfy the required physical constraints.  We compare the results with those obtained using conventional ML-QEM, ZNE and readout error mitigation (REM). Ground-state energies and quantum-state fidelities are used to evaluate the performance of these different error-mitigation methods. 

The remainder of this paper is organized as follows. Section~\ref{2} describes the three-qubit $\Lambda$-deuteron and four-qubit $\Lambda$-$\alpha$ mapped Hamiltonians, their implementation on the IBM Quantum platform, and the baseline results obtained using ZNE and REM. Section~\ref{3} presents the proposed ML-QEM framework, including the training-data strategies, regression models, and convex projection procedure. Section~\ref{4} presents the results and discussion for both hypernuclear systems, including the ground-state energies, quantum-state fidelities, and performance of the ML models. Finally, Sec.~\ref{5} summarizes the main findings and discusses possible future research directions. Additional data and metrics are provided in the Appendix.

\section{\label{2} Quantum simulation of the Ground-state energies of Hypernuclear systems}

We start by writing the Hamiltonian for bound hypernuclear systems, viz., the  $\Lambda$-deuteron ($^{3}\mathrm{H}_\Lambda$) and $\Lambda$-$\alpha$ ($^{5}\mathrm{He}_\Lambda$), expressed in harmonic oscillator (HO) basis as:
\begin{equation}
	H_N = \sum_{i,j=0}^{N-1} h_{ij} a_i^\dagger a_j,
	\label{Hamiltonian}
\end{equation}
where $h_{ij}$ are the Hamiltonian matrix elements given by 
\begin{eqnarray}
	h_{ij} &=& \langle i | \hat{T} + \hat{U}_{{}} | j \rangle\nonumber\\
	&=&  \int \phi_i^*(r) 
	\left[ -\frac{\hbar^2}{2m} \nabla^2 +U_{}(r) \right] \phi_j(r) \, dr.
	\label{matrix}
\end{eqnarray}
Here, $|i\rangle$ and $|j\rangle$ denote the HO basis states with coordinate-space wavefunctions $\phi_i(r)$ and $\phi_j(r)$, respectively. The matrix elements $h_{ij}$ are calculated using the spherical HO basis, whose radial wavefunctions are given by: 
\begin{equation}
	\begin{aligned}
		R_{nl}(r)=
		\sqrt{\frac{2\,n!}{b^{3}\,\Gamma\!\left(n + l + \tfrac{3}{2}\right)}}
		x^{\!l}
		\exp\!\left[-\frac{x^{2}}{2}\right]
		L_{n}^{\,l + 1/2}\!\left(x^2\right)
        \label{HO}
	\end{aligned}
\end{equation}
with $x=\frac{r}{b}$, $\Gamma()$  is the Gamma function, and $L_n^{l}()$ is the generalized Laguerre polynomial \cite{arfken2011mathematical}. The HO basis is characterized by the length parameter $b=\sqrt{\hbar/(m \omega)}$, where $m$ denotes the reduced mass of the system, and $\omega$ is the  frequency of the oscillator. 

The kinetic energy and potential energy matrix elements of equation~(\ref{matrix}) are determined as:
\begin{equation}
	\begin{split}
		\langle n' l'| T | n l \rangle 
		&= 
		\biggl[
		\left(2n + l + \tfrac{3}{2}\right)\delta_{n}^{n'}
		+ \sqrt{n\!\left(n + l + \tfrac{1}{2}\right)}\,\delta_{n}^{\,n'+1} \\
		&\quad
		+ \sqrt{(n + 1)\!\left(n + l + \tfrac{3}{2}\right)}\,\delta_{n}^{\,n'-1}
		\biggr]\delta_{l}^{l'}\,\frac{\hbar\omega}{2},\nonumber 
	\end{split}
\end{equation}

\begin{eqnarray}
	\langle n' l'| U_{\Lambda d} | n l \rangle = \delta_l^{l'} \int R_{n' l'}(r) \,
	U_{\Lambda d}(r) \, r^2 \, R_{nl}(r) \, dr.\nonumber \\
	\
\end{eqnarray}

To simulate the ground state energies of the considered hypernuclei on a quantum computer, we need to map the fermionic operators of the Hamiltonian in equation (\ref{Hamiltonian}) onto qubit operators. Several fermion-to-qubit mappings have been introduced in the literature \cite{mcardle2020quantum,di2021improving, seeley2012bravyi,somma2002simulating, bravyi2002fermionic,jordan1928paulische}. In this work,  we use the Jordan-Wigner (JW) transformation \cite{jordan1928paulische,siwach2021quantum}, which maps the fermionic creation and annihilation operators in equation (\ref{Hamiltonian}) onto qubits as follows:
\begin{align}
	a_n^\dagger &\rightarrow \frac{1}{2} \left( \prod_{j=0}^{n-1} \sigma_z^j \right)(\sigma_x^n - i\,\sigma_y^n), \nonumber \\
	a_n &\rightarrow \frac{1}{2} \left( \prod_{j=0}^{n-1} \sigma_z^j \right)(\sigma_x^n + i\,\sigma_y^n).
	\label{jordon}
\end{align}

Here, \(\sigma_x^n\), \(\sigma_y^n\), and \(\sigma_z^n\) denote the Pauli-\(X\), Pauli-\(Y\), and Pauli-\(Z\) operators, respectively, acting on the \(n\)-th qubit.

\subsection{Three-qubit mapped Hamiltonian of $\Lambda$-deuteron system}

To model the bound $\Lambda$-deuteron ($^{3}\mathrm{H}_{\Lambda}$) system, we treat the deuteron as an inert core interacting with $\Lambda$ hyperon through the Woods-Saxon potential proposed by Bertulani \cite{bertulani2023probing},
\begin{equation}
	U_{\Lambda d}(r)=\frac{-U_0}{1+\exp\left(\frac{r-R}{a}\right)},
\end{equation}
where $U_0$, $R$, and $a$ denote the potential depth, radius, and diffuseness, respectively. Throughout this work, we use $U_0=17.16$ MeV, $R=2.5$ fm, and $a=0.65$ fm. The experimental binding energy of the $\Lambda$-deuteron system is $B_{\Lambda d}=-0.13\pm0.05~\mathrm{MeV}$, reported in~\cite{jurivc1973new}. For the length parameter $b$, the harmonic-oscillator frequency is determined by adopting $R_{\Lambda d}\approx10$ fm following Ref.~\cite{bertulani2023probing}. 

Applying the Jordan-Wigner transformation to Eq.~(\ref{Hamiltonian}) along with above potential yields the following three-qubit Hamiltonian for the $\Lambda$-deuteron system as 
\begin{align}
	H_3 &= 1.222I - 0.043 \sigma_z^0 
	- 0.046 \left(\sigma_x^0 \sigma_x^1 + \sigma_y^0 \sigma_y^1 \right) \nonumber \\
	&\quad - 0.391 \sigma_z^1 
	- 0.308 \left(\sigma_x^0 \sigma_z^1 \sigma_x^2 
	+\sigma_y^0 \sigma_z^1 \sigma_y^2 \right) \nonumber \\
	&\quad - 0.787 \sigma_z^2 
	+ 0.118 \left(\sigma_x^1 \sigma_x^2 + \sigma_y^1 \sigma_y^2 \right).
	\label{JW_transformed}
\end{align}

Exact diagonalization of Eq.~(\ref{JW_transformed}) yields a ground-state energy of $-0.135$ MeV, which serves as the reference value for further study.

\subsection{ Four-qubit mapped Hamiltonian of $\Lambda-\alpha$ system}
The interaction between bound $\Lambda$-$\alpha$ ($^{5}\mathrm{He}_{\Lambda}$) system is described by using two phenomenological Gaussian potentials, namely the Gibson~I and Bassichis-Gal \textit{c} models \cite{daskaloyannis1982determination,portilho19919be},
\begin{equation}
	U_{\Lambda\alpha}(r)=\sum_i U_i\exp\left[-\left(\frac{r}{a_i}\right)^2\right],
\end{equation}
where $U_i$ and $a_i$ denotes the potential strength and range parameter, respectively. The Gibson~I interaction is represented by a single attractive Gaussian potential, whereas the Bassichis-Gal \textit{c} interaction consists of a two-range Gaussian potential with a long-range attractive component and a short-range repulsive core. The experimental binding energy of the $\Lambda$-$\alpha$ system is $B_{\Lambda \alpha}=-3.10\pm0.02~\mathrm{MeV}$, reported in Ref.~\cite{jurivc1973new}. The adopted parameters are summarized in Table~\ref{tab:potential_parameters_4qubit}. 

\begin{table}[t]
	\small
	\caption{\label{tab:potential_parameters_4qubit}
		Potential parameters for the Gibson I and Bassichis-Gal \textit{c}  interaction models for $\Lambda-\alpha$ system. Potential strengths are given in MeV and range parameters in fm.}
	\begin{ruledtabular}
		\begin{tabular}{lcc}
			Parameter  & Gibson I    & Bassichis-Gal \textit{c} \\
			\hline
			$U_1$  & $-44.82$  & $-52.78$ \\
			$U_2$  & ---  & $9.99$ \\
			$a_1$  & $1.576$  & $1.564$ \\
			$a_2$ & ---  & $1.316$ \\
			$R_{\Lambda\alpha}$  & $2.78$  & $2.79$ \\
		\end{tabular}
	\end{ruledtabular}
\end{table}

Applying the Jordan-Wigner transformation to Eq.~(\ref{Hamiltonian}) yields the following four-qubit Hamiltonians corresponding to the Gibson I and Bassichis-Gal \textit{c} interactions, respectively

\begin{eqnarray}
	H_4^G &=& 27.915I+0.879 \sigma_z^0 - 4.549 \sigma_z^1 - 9.679 \sigma_z^2\nonumber\\&-& 0.733(\sigma_x^0\sigma_x^1+\sigma_y^0\sigma_y^1) + 2.193(\sigma_x^1\sigma_x^2+\sigma_y^1\sigma_y^2)\nonumber\\&-& 2.612(\sigma_x^0\sigma_z^1\sigma_x^2+\sigma_y^0\sigma_z^1\sigma_y^2) - 14.566 \sigma_z^3\nonumber\\ &-& 1.890(\sigma_x^0 \sigma_z^1 \sigma_z^2\sigma_x^3 + \sigma_y^0 \sigma_z^1 \sigma_z^2 \sigma_y^3) \nonumber\\&-& 2.304(\sigma_x^1 \sigma_z^2 \sigma_x^3 + \sigma_y^1 \sigma_z^2 \sigma_y^3)\nonumber\\ &+& 4.891(\sigma_x^2\sigma_x^3+\sigma_y^2\sigma_y^3),
	\label{transformed1}
\end{eqnarray}

\begin{eqnarray}
	H_4^B &=& 27.992I+0.908 \sigma_z^0- 4.576 \sigma_z^1- 9.719 \sigma_z^2\nonumber\\&-& 0.720(\sigma_x^0\sigma_x^1+\sigma_y^0\sigma_y^1) + 2.233(\sigma_x^1\sigma_x^2+\sigma_y^1\sigma_y^2)\nonumber\\&-& 2.569(\sigma_x^0\sigma_z^1\sigma_x^2+\sigma_y^0\sigma_z^1\sigma_y^2)-14.605 \sigma_z^3\nonumber\\ &-& 1.831(\sigma_x^0 \sigma_z^1 \sigma_z^2\sigma_x^3 + \sigma_y^0 \sigma_z^1 \sigma_z^2 \sigma_y^3) \nonumber\\&-& 2.255(\sigma_x^1 \sigma_z^2 \sigma_x^3 + \sigma_y^1 \sigma_z^2 \sigma_y^3)\nonumber\\ &+& 4.933(\sigma_x^2\sigma_x^3+\sigma_y^2\sigma_y^3).
	\label{transformed2}
\end{eqnarray}

Exact diagonalization of Eqs.~(\ref{transformed1}) and (\ref{transformed2}) yields ground-state energies of  $-3.118$ MeV and $-3.115$ MeV for the Gibson I and Bassichis-Gal \textit{c} interactions, respectively. 

\subsection{Experimental Execution and Baseline Error Mitigation}
After mapping the hypernuclear Hamiltonians onto qubit operators, the ground-state energies are estimated using the Variational Quantum Eigensolver (VQE) \cite{peruzzo2014variational,mcclean2016theory,wecker2015progress}, a widely used quantum-classical hybrid algorithm. Within the variational framework, a parameterized trial state $|\psi(\theta)\rangle$ is optimized by minimizing the expectation value of a Hamiltonian operator $\hat{H}$  
\begin{equation}
	E(\theta)=\langle\psi(\theta)|\hat{H}|\psi(\theta)\rangle.
\end{equation}

The variational ans\"atze for the three- and four-qubit Hamiltonians follow the circuit structures proposed in Ref.~\cite{shehab2019toward} and are shown in Fig.~\ref{fig:circuits}.

\definecolor{rygateTop}{HTML}{DCE8F7}   
\definecolor{rygateBot}{HTML}{A9C6E8}   
\definecolor{xgateTop}{HTML}{FDEAD9}    
\definecolor{xgateBot}{HTML}{F0B48A}    
\definecolor{measure}{HTML}{EDEDED}     

\tikzset{
  ry gate/.style={top color=rygateTop,bottom color=rygateBot,draw=black!70},
  x gate/.style={top color=xgateTop,bottom color=xgateBot,draw=black!70},
  meas/.style={fill=measure,draw=black!70},
}

\newcommand{\circfont}{\small}   
\newcommand{\ql}[1]{\lstick{\circfont $q_{#1}$}}   
\newsavebox{\circB}

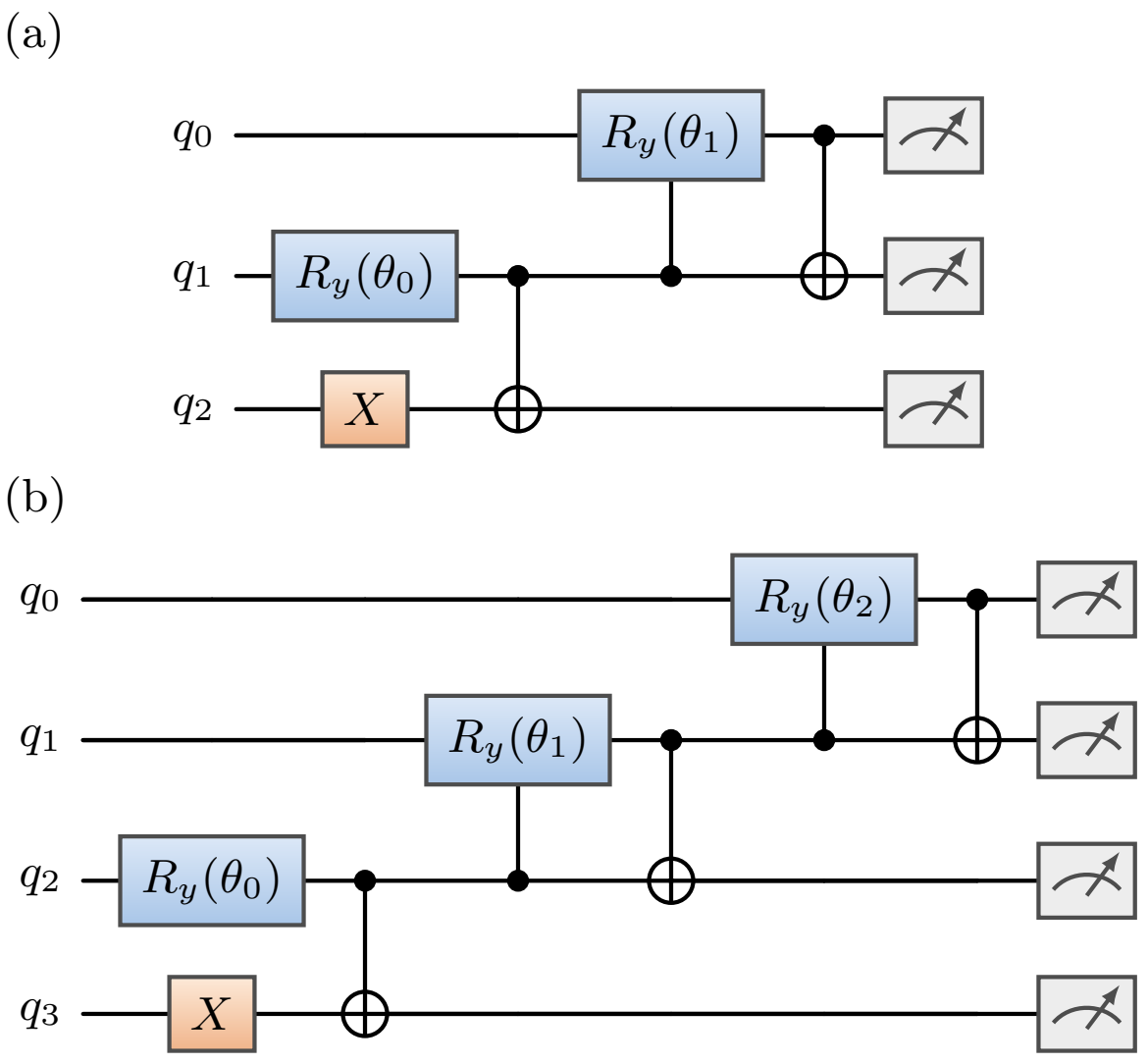
\begin{figure}[t]
	\centering
	\sbox{\circB}{%
		\begin{quantikz}[font=\circfont, column sep=0.25cm, row sep=0.35cm]
			\lstick{$q_0$} & \qw & \qw & \qw & \qw & \gate[style={ry gate}]{R_y(\theta_2)} & \ctrl{1} & \meter[style={meas}]{} \\
			\lstick{$q_1$} & \qw & \qw & \gate[style={ry gate}]{R_y(\theta_1)} & \ctrl{1} & \ctrl{-1} & \targ{} & \meter[style={meas}]{} \\
			\lstick{$q_2$} & \gate[style={ry gate}]{R_y(\theta_0)} & \ctrl{1} & \ctrl{-1} & \targ{} & \qw & \qw & \meter[style={meas}]{} \\
			\lstick{$q_3$} & \gate[style={x gate}]{X} & \targ{} & \qw & \qw & \qw & \qw & \meter[style={meas}]{}
		\end{quantikz}%
	}
	\typeout{CIRCUIT B WIDTH = \the\wd\circB, COLUMNWIDTH = \the\columnwidth}

		\begin{tabular}{@{}c@{}}
		\multicolumn{1}{@{}l@{}}{\small (a)}\\
		\begin{quantikz}[font=\circfont, column sep=0.25cm, row sep=0.35cm]
			\lstick{$q_0$} & \qw & \qw & \gate[style={ry gate}]{R_y(\theta_1)} & \ctrl{1} & \meter[style={meas}]{} \\
			\lstick{$q_1$} & \gate[style={ry gate}]{R_y(\theta_0)} & \ctrl{1} & \ctrl{-1} & \targ{} & \meter[style={meas}]{} \\
			\lstick{$q_2$} & \gate[style={x gate}]{X} & \targ{} & \qw & \qw & \meter[style={meas}]{}
		\end{quantikz}
		\\[18pt]
		\multicolumn{1}{@{}l@{}}{\small (b)}\\
		\usebox{\circB}
	\end{tabular}

	\caption{Variational quantum circuits for the (a) three-qubit $\Lambda$-deuteron and (b) four-qubit $\Lambda$-$\alpha$ systems. The circuits prepare the trial state using a $X$ gate, followed by parameterized $R_y$ rotations and CNOT entangling operations.}
	\label{fig:circuits}
\end{figure}

Since hardware optimization requires multiple circuit evaluations, making direct cloud-based execution through VQE very challenging. So, the optimal variational parameters $\theta_i^{\mathrm{opt}}$ were first determined using classical VQE simulations and subsequently implemented on the IBM quantum processor i.e., \textit{ibm\_kingston}. The optimized parameters for the both three- and four-qubit Hamiltonians are listed in Table~\ref{tab:3qubit_optimized_theta_deg} and Table~\ref{tab:optimized_theta_deg_4qubit}.

\begin{table}[h!]
	\begin{ruledtabular}
		\caption{
			Classically optimized variational parameters $\theta_i^{\text{opt}}$ (in degrees) 
			for the three-qubit $\Lambda$-deuteron system.
		}
		\label{tab:3qubit_optimized_theta_deg}
		\begin{tabular}{ccc}
			\text{Potential}  & \text{Parameter} ($\theta_i$) & $\theta_i^{\text{opt}}$ \text{[deg]} \\
			
			\hline
			\multirow{2}{*}{\centering Woods-Saxon} 
			& \( \theta_0 \) & 140.49\\
			
			& \( \theta_1 \) & 179.06 \\
		\end{tabular}
	\end{ruledtabular}
\end{table}

\begin{table}[t]
	\begin{ruledtabular}
		\caption{Classically optimized variational parameters $\theta_i^{\mathrm{opt}}$ (in degrees) for Gibson~I and Bassichis-Gal \textit{c} potentials of four-qubit $\Lambda$-$\alpha$ system.}
		\label{tab:optimized_theta_deg_4qubit}
		\begin{tabular}{lcc}
			Potential & Parameter ($\theta_i$) & $\theta_i^{\mathrm{opt}}$ (deg) \\
			\hline
			\multirow{3}{*}{Gibson~I}
			& $\theta_0$ & 171.85 \\
			& $\theta_1$ & 159.10 \\
			& $\theta_2$ & 170.73 \\
			\hline
			\multirow{3}{*}{Bassichis-Gal \textit{c}}
			& $\theta_0$ & 172.38 \\
			& $\theta_1$ & 159.30 \\
			& $\theta_2$ & 171.35 \\
		\end{tabular}
	\end{ruledtabular}
\end{table}

\begin{figure}[t]
	\centering
	\captionsetup{}
	\includegraphics[width=\columnwidth]{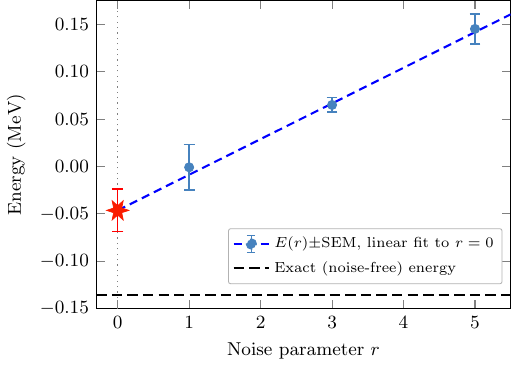}
	\caption{Linear fit of the binding energy (MeV) for $\Lambda$-deuteron system as a function of the noise parameter $r$. Red star of the fitted line corresponds to the extrapolated noiseless estimate of the ground-state energy with error bars from the fit uncertainty, as summarized in Table~\ref{3qubit_energy_comparison}. The black dashed line is the exact noise-free energy $-0.135$~MeV.}
	\label{fig_3qubit_ZNE}
\end{figure}

The ground-state energies of the bound systems are computed using IBM's Qiskit software stack \cite{aleksandrowicz2019qiskit}, an open-source, Python-based framework for quantum computing. Qiskit provides access to several computational backends, ranging from idealized simulators, such as the state-vector simulator, to real superconducting quantum processors, including \textit{ibm\_kingston}. Calculations were first performed using the state-vector simulator, and the resulting energies are reported in Table~\ref{3qubit_energy_comparison} and Table~\ref{4qubit_energy_comparison} for both the three- and four-qubit Hamiltonians. The energies obtained from the statevector simulator are in exact agreement with the energy eigenvalues obtained by exact diagonalization of the mapped Hamiltonians, confirming the validation of  both Hamiltonians construction and the implementation of the variational circuits.

The variational circuits were then executed on the 156-qubit \textit{ibm\_kingston} processor which returns measurement outcomes in a little-endian format (the rightmost bit in the bit string corresponds to qubit 0). To ensure the correct evaluation of expectation values, qubit ordering of each Pauli string is chosen consistently with this convention. Expectation values were evaluated with the Qiskit Runtime \texttt{EstimatorV2} primitive at 10,000 shots, which groups commuting Pauli terms and applies the required measurement-basis rotations internally. Although the Hamiltonians and VQE circuits described above provide a framework for simulating these hypernuclear systems, noise in current quantum hardware can reduce the accuracy of the measured energies. 

\begin{table}[h!]
	\begin{ruledtabular}
		\caption{Ground-state energies (in MeV) obtained from the state vector simulations, IBM raw, REM corrected and ZNE with REM techniques for the three-qubit Hamiltonian.}
		\label{3qubit_energy_comparison}
		\begin{tabular}{c  cc cc}
			Potential    &	Ideal & Raw &REM&ZNE +REM \\
			\hline
			\multirow{2}{*}{Woods-Saxon} & \multirow{2}{*}{$-0.135$} & \multirow{2}{*}{$0.618 $}&$-0.001$ & $-0.046 $\\
			& & &$\pm0.024$ & $ \pm 0.022$\\
		\end{tabular}
	\end{ruledtabular}
\end{table}

\begin{figure}
	\centering
	\captionsetup{}
	\includegraphics[width=\columnwidth]{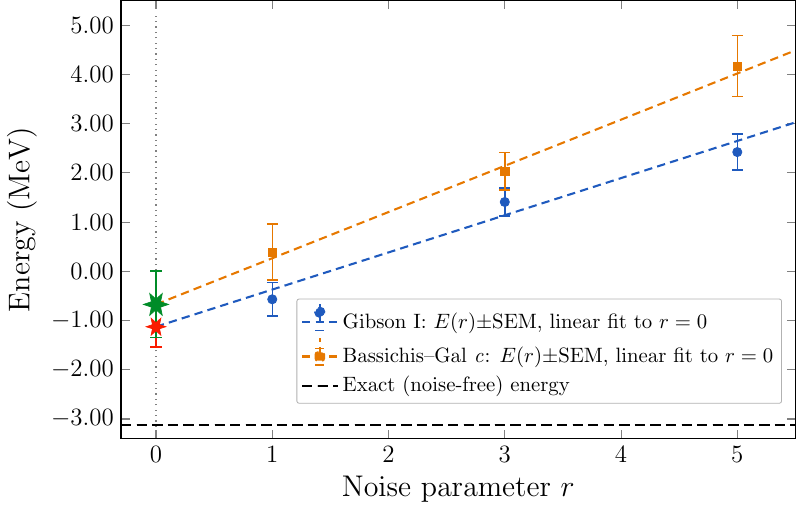}
	\caption{Linear fit of the binding energy (MeV) for $\Lambda$-$\alpha$ system as a function of the noise parameter $r$. Red and green stars of the fitted lines corresponds to the extrapolated noiseless estimate of the ground-state energies for Gibson~I and Bassichis-Gal \textit{c} potentials respectively, as summarized in Table~\ref{4qubit_energy_comparison}. The black dashed line is the exact noise-free energy $\approx -3.11$~MeV for both potentials.}
	\label{fig_4qubit_ZNE}
\end{figure}

\begin{table}[h!]
	\begin{ruledtabular}
		\caption{Ground-state energies (in MeV) obtained from the state vector simulations, noisy IBM quantum hardware, REM corrected and ZNE with REM for the four-qubit system, using the Gibson~I and Bassichis-Gal \textit{c} nuclear potentials.}
		\label{4qubit_energy_comparison}
		\begin{tabular}{c c c c c}
			Potential    &	Ideal & Raw &REM&ZNE +REM \\
			\hline
			\multirow{2}{*}{Gibson~I}  & \multirow{2}{*}{$-3.118$} & \multirow{2}{*}{$6.851 $ }& $-0.568$& $-1.125 $\\
			&  &  & $\pm0.343$& $ \pm 0.412$\\
			\hline
			\multirow{2}{*}{Bassichis-Gal \textit{c}} & \multirow{2}{*}{$-3.115$} & \multirow{2}{*}{$6.569 $} & $0.391$& $-0.672 $\\
			& &   & $\pm0.570$& $ \pm 0.676$\\
		\end{tabular}
	\end{ruledtabular}
\end{table}

As baseline error mitigation techniques, readout error mitigation (REM) and zero-noise extrapolation (ZNE) were applied. REM was enabled through resilience level 1 of the \texttt{EstimatorV2} primitive. ZNE is achieved by repeating  the whole unitary with noise parameter $r = 1,3,5, \ldots$ \cite{temme2017error,giurgica2020digital} and then extrapolating the expectation values of the corresponding Hamiltonians to the noiseless limit. For each $r$, five independent hardware runs were collected with REM enabled. 

The physical qubit layout was selected once for the three- and four-qubit circuits and held fixed throughout the noise-scaling procedure, so that variations in hardware noise do not affect the extrapolation procedure. All circuits were transpiled at optimization level~0 with this fixed layout, which prevents the compiler from cancelling the inserted $U^\dagger U$ pairs used for noise scaling. The mean energy and its standard error of the mean (SEM) were computed for each $r$, and the energies were extrapolated to $r=0$ using a weighted least-squares linear fit using inverse-variance weights ($1/\mathrm{SEM}^2$). The uncertainty in the extrapolated energy estimated from the covariance matrix of the fit. As shown in Fig.~\ref{fig_3qubit_ZNE} and Fig.~\ref{fig_4qubit_ZNE}, the IBM  output as a function of the noise parameter $r$ is well approximated by straight lines and intercepts of these lines at $r=0$ are listed in Table~\ref{3qubit_energy_comparison} and Table~\ref{4qubit_energy_comparison} for both three- and four-qubit systems.

Although REM and ZNE reduce the effects of hardware noise, noticeable discrepancies from the theoretical ground-state energies remain. These remaining errors motivates the machine learning based error mitigation approach introduced in the next section.

\begin{figure*}
	\centering
	\captionsetup{}
	\includegraphics[width=\textwidth]{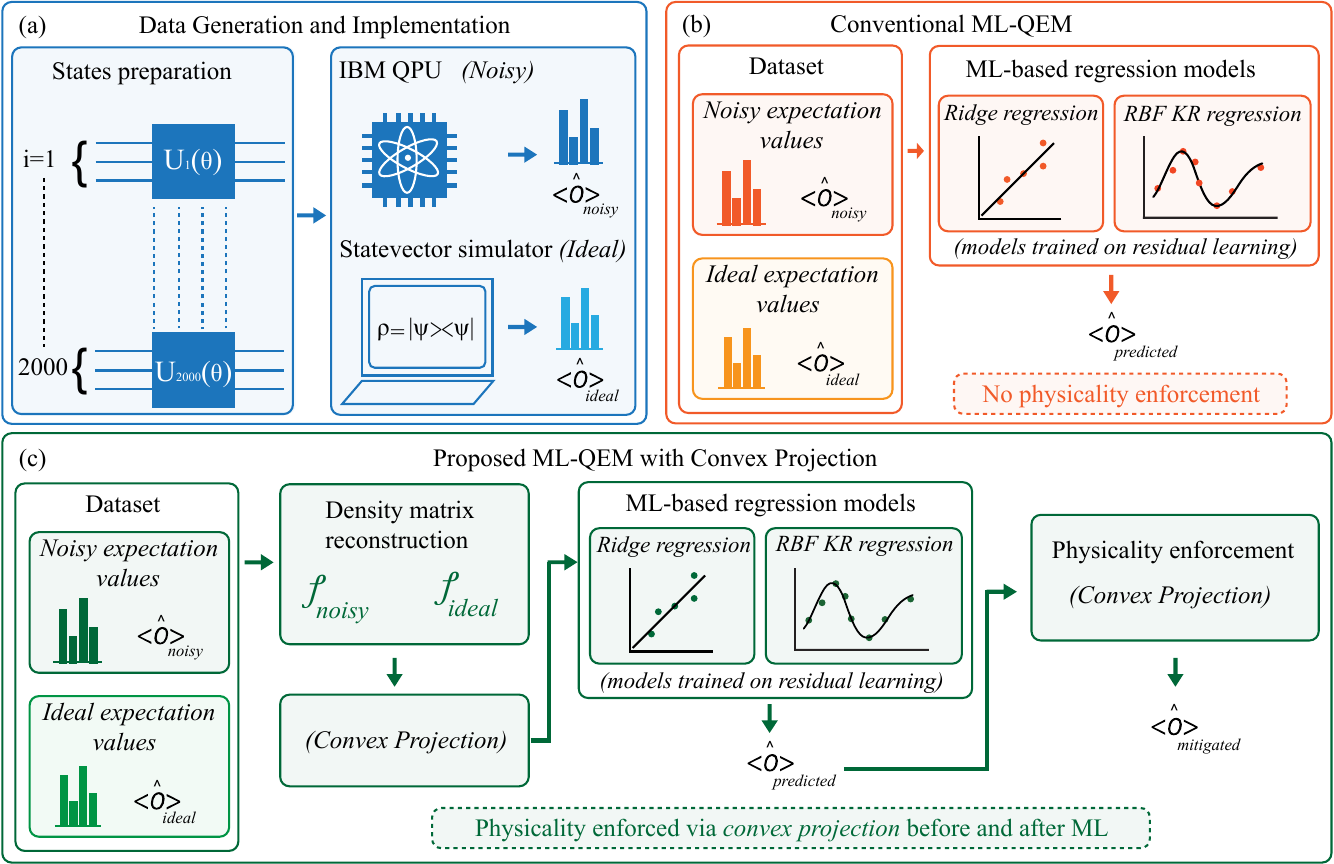}
	\caption{Workflow of the conventional and proposed ML-QEM framework. \textbf{(a)} A dataset of 2000 three-qubit quantum states (Haar-random and VQE-ansatz states) is run on an IBM Quantum processor and an ideal statevector simulator to obtain  noisy and ideal Pauli expectation values, respectively. \textbf{(b)} In conventional ML-QEM, a regression model is trained on the noisy data to predict the correction, which is then added to the noisy values to give the mitigated values. \textbf{(c)} In the proposed ML-QEM framework, the noisy and ideal data are first converted to density matrices and checked for physical validity. Any invalid state is replaced by the nearest valid state using convex projection, and then corrected data are used to train the model. The model predictions are again checked and corrected in the same way to give the final mitigated values. The same procedure is followed for the four-qubit case.}
	\label{fig:workflow}
\end{figure*}

\section{\label{3}Machine Learning-Based Quantum Error Mitigation Framework}

The overall workflow of the conventional and proposed machine learning-based quantum error mitigation (ML-QEM) frameworks is shown in Fig.~\ref{fig:workflow}. The procedure consists of three main steps. First, we generate training datasets using Haar-random and variational quantum eigensolver (VQE) ansatz states. The noisy Pauli expectation values are obtained from an IBM Quantum processor, while the corresponding ideal values are calculated using a statevector simulator as shown in Fig.~\ref{fig:workflow}(a). Next, in the conventional ML-QEM approach, regression models are trained using the noisy and ideal expectation values to predict measurement errors as shown in the Fig.~\ref{fig:workflow}(b). Finally, in the proposed approach, convex projection is applied to the reconstructed density matrices before and after machine learning to ensure that the quantum states satisfy the physical constraints as shown in the Fig.~\ref{fig:workflow}(c). The following subsections describe the training-data generation, regression models, convex projection method, and evaluation metrics.

\subsection{Training data and Implementation}

For dataset generation, we consider two types of training states. The first dataset contains Haar-random quantum states, and the second contains VQE-ansatz states generated using the variational circuits employed in this work. Each dataset contains 2000 states separately.

Haar-random quantum states are pure states from the Haar measure using the \texttt{random\_statevector} function of Qiskit \cite{zyczkowski2001induced}. These states are spread evenly over the whole Hilbert space, making them general training dataset that is not restricted to any specific circuit class. On the quantum processor, each state was prepared with the \texttt{initialize} routine of Qiskit, which makes relatively deep state-preparation circuits \cite{iten2016quantum}. 

The VQE-ansatz states are generated using the variational circuits that we use for each Hamiltonians. These states belong to a small family of states that includes the variational ground state.  The ansatz covers the three-dimensional single-excitation subspace with real amplitudes. For each state, we drew every angle independently and uniformly from $[0,2\pi)$. Uniform sampling of the angles is not the same as Haar sampling, so some regions of the state family are sampled more often than others. The parameters and state vectors were generated once and saved, so all subsequent steps use the same states.

For every state, we computed the expectation values of all $4^{n}$ Pauli strings, where $n$ is the number of qubits, yielding $64$ strings for $n=3$ and $256$ strings for $n=4$. The ideal Pauli expectation values were computed exactly from the statevector simulator, whereas the noisy expectation values were measured on the IBM Quantum processor \textit{ibm\_kingston} using the Qiskit Runtime \texttt{EstimatorV2} primitive with 1024 shots per circuit. We used default IBM readout-error mitigation (TREX, resilience level~1) without applying any additional error-mitigation techniques. The circuits were transpiled with the Qiskit preset pass manager at optimization level~1, and the observables were mapped to the physical qubits chosen by the transpiler.

Finally, we test our proposed ML-QEM framework on the three- and four-qubit mapped Hamiltonians describing the $\Lambda$-deuteron and $\Lambda$-$\alpha$ systems, respectively. These target states were not part of any training dataset, and executed once on \textit{ibm\_kingston} using the same settings described above. 

\subsection{Residual Prediction and Learning Models}
\label{RR}
Each quantum state is represented by the vector of its non-identity Pauli expectation values. Let $\mathbf{x}\in\mathbb{R}^{d}$ denote the noisy Pauli expectation values measured on the IBM processor and $\mathbf{y}\in\mathbb{R}^{d}$ the corresponding ideal values, with $d=4^n-1$. Instead of predicting $\mathbf{y}$ directly, a regressor $f$ is trained to predict the residual. We define the residual $\mathbf{r}$ as the difference between the ideal and noisy expectation values.
\begin{equation}
	\mathbf{r}=\mathbf{y}-\mathbf{x}.
\end{equation}
For a given noisy input $\mathbf{x}$, the regressor predicts the residual as $\mathbf{r}_{\mathrm{predicted}}=f(\mathbf{x})\approx\mathbf{r}$. The mitigated expectation values are then obtained as
\begin{equation}
	\hat{\mathbf{y}}=\mathbf{x}+\mathbf{r}_{\mathrm{predicted}}.
\end{equation}

We use two regression models: ridge regression and radial basis function-based kernel ridge regression (RBF-KRR) in our study \cite{murphy2012machine,mcdonald2009ridge,saunders1998ridge}. Ridge regression is a linear regression model, used to map the linear relations with an $L_2$ regularization strength parameter set at $\alpha=1$. In contrast, RBF-KRR with a Gaussian kernel, captures the nonlinear relationships between noisy and ideal expectation values. RBF-KRR uses the same $L_2$ regularization strength at $\alpha_{\mathrm{K}}=0.1$, while the kernel width $\gamma$ is set to the inverse of the median pairwise squared distance between input vectors.

Each dataset contains $2000$ states, which are divided into $200$ validation states, $200$ test states, and a training pool of $1600$ states. In our datasets, all $2000$ ideal states are distinct. To study the effect of training data size, we train the models using $N\in\{50,100,200,400,800,1600\}$ states from the training pool and repeat the procedure for $10$ different seeds. The same trained model are applied to the held-out test states and to the final target states, which are never used for training. All hyperparameters were fixed a priori and kept identical across training sizes and random seeds.

The predictive performance of the models on the test states is evaluated using three metrics, computed between the predicted and ideal Pauli expectation values: the mean absolute error (MAE), the root mean square error (RMSE), and the coefficient of determination ($R^2$) \cite{willmott2005advantages, chicco2021coefficient}. They are defined as
\begin{align}
    \mathrm{MAE} &= \frac{1}{M}\sum_{i=1}^{M}|y_i-\hat{y}_i|,
    \label{eq:MAE} \\[6pt]
    \mathrm{RMSE} &= \sqrt{\frac{1}{M}\sum_{i=1}^{M}(y_i-\hat{y}_i)^2},
    \label{eq:RMSE} \\[6pt]
    R^2 &= 1-\frac{\sum_{i=1}^{M} (y_i-\hat{y}_i)^2}
    {\sum_{i=1}^{M} (y_i-\bar{y})^2}.
    \label{eq:R2}
\end{align}
where $M$ is the total number of Pauli expectation values over all test states, $y_i$ and $\hat{y}_i$ are the ideal and predicted values of the $i$-th Pauli expectation value, and $\bar{y}$ is the mean of the ideal values. The same metrics evaluated on the noisy values are also reported for reference.

\subsection{Convex projection onto Physical states}

As described in the previous subsection~\ref{RR}, each quantum state is represented by a vector of non-identity Pauli expectation values. These expectation values can be used to reconstruct the corresponding density matrix. Any $n$-qubit quantum state can be written in terms of its Pauli expectation values as
\begin{equation}
	\rho = \frac{1}{2^n}\sum_{i=1}^{4^n} c_i P_i ,
	\label{eq:rho}
\end{equation}
where $c_i = \mathrm{Tr}(\rho P_i)$ is the expectation value of the Pauli operator $P_i$. A valid quantum state must be Hermitian, have unit trace, and be positive semidefinite. The operator in Eq.~(\ref{eq:rho}) is Hermitian and has unit trace since the coefficients $c_i$ are real and the identity Pauli operator has an expectation value of one. However, it generally violates positive semidefiniteness because of experimental noise, finite measurement statistics and reconstruction errors \cite{singh2016constructing}. 

In our experiments, the noisy Pauli expectation-value vectors were reconstructed using quantum state tomography (QST)~\cite{aleksandrowicz2019qiskit}. For both $n=3$ and $n=4$, all 2000 reconstructed noisy vectors exhibited at least one negative eigenvalue. This shows that the reconstructed operators do not represent physically valid quantum states.

To obtain physically valid quantum states, we project each reconstructed operator onto the set of physical density matrices by minimizing the squared Frobenius distance. We denote the set of valid  $n$-qubit density matrices by $\mathcal{D}_n$, defined as follows: 
\begin{equation}
\mathcal{D}_n = \left\{
\sigma \in \mathbb{C}^{2^n \times 2^n}
\;\middle|\;
\begin{aligned}
\sigma &= \sigma^\dagger,\\
\mathrm{Tr}(\sigma) &= 1,\\
\sigma &\succeq 0
\end{aligned}
\right\}.
\end{equation}

The physical state closest to the reconstructed operator $\rho$ is obtained by solving
\begin{equation}
\rho_{\mathrm{phys}} =
\arg\min_{\sigma \in \mathcal{D}_n}
\left\| \sigma - \rho \right\|_F^2 ,
\label{eq:proj}
\end{equation}
where $\sigma$ is the optimization variable and $\|\cdot\|_F$ denotes the Frobenius norm. Since $\mathcal{D}_n$ is a closed and convex set, the solution of this optimization problem is unique ~\cite{smolin2012efficient,wang2013projection}.

To perform the projection, we first diagonalize the reconstructed operator $\rho$. The eigenvectors remain unchanged during the projection, while the eigenvalues are replaced by the closest non-negative values that sum to one. The projected density matrix therefore satisfies all the physicality conditions.  The projection also leaves already valid physical states unchanged and does not increase the Frobenius distance between the reconstructed operator and any physically valid density matrix.

We apply convex projection at two stages to study its effect on ML-QEM performance. First, the experimentally reconstructed noisy input vectors are projected onto the set of valid quantum states before training the ML model. Second, the corrected Pauli expectation values predicted by the ML model are used to reconstruct a density matrix, which is then projected onto the same physical state space. In this way, convex projection is applied as a pre-processing and post-processing step, while the ML regressor itself is trained without any explicit physicality constraints.

\subsection{Evaluation metrics}

To assess the effectiveness of the proposed framework, we use two quantities for the $\Lambda$-deuteron and $\Lambda$-$\alpha$ target systems: ground-state energy and state fidelity. The Hamiltonian is expressed in terms of Pauli strings as $H=\sum_i h_iP_i$, where $P_i$ are Pauli operators and $h_i$ are real coefficients. For a set of Pauli expectation values $c_i$, the corresponding energy is calculated as
\begin{equation}
E=\sum_i h_i c_i .
\end{equation}

We use the Uhlmann-Jozsa fidelity to compare the reconstructed states with the corresponding ideal target states~\cite{uhlmann1975transition,jozsa1994fidelity}. For a reconstructed density matrix $\rho$ and an ideal target density matrix $\sigma^*$, the fidelity is defined as
\begin{equation}
F(\rho,\sigma^*) =
\left(\operatorname{Tr}\sqrt{\sqrt{\rho}\,\sigma^*\,\sqrt{\rho}}\right)^2 .
\label{eq:fidelity}
\end{equation}

Since the fidelity measures the closeness of two quantum states and requires both density matrices to be physically valid. Therefore, we project the outputs of the unprojected pipelines onto the set of valid density matrices before calculating the fidelity. The energy, however, is calculated directly from the raw Pauli expectation values, without applying physical-state projection. This gives the energy estimates directly from the ML predictions, without any additional correction due to projection.

\section{\label{4}Results and Discussion}

\begin{figure*}[t]
	\includegraphics[width=1\textwidth]{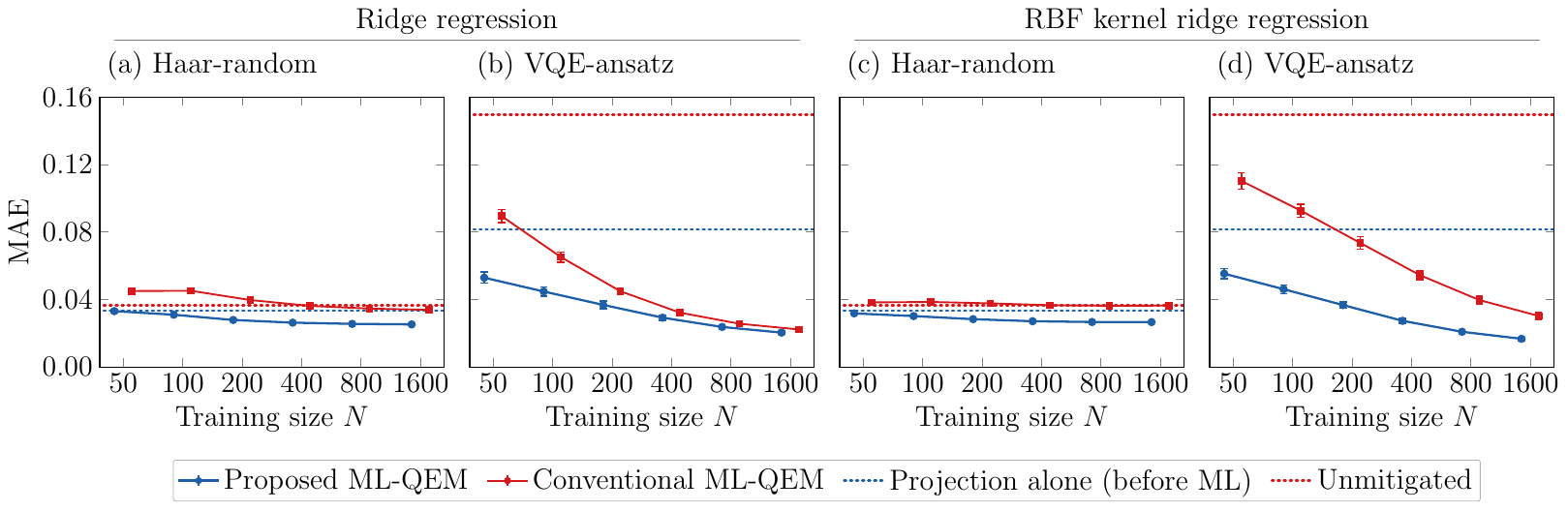}
       \caption{Test-set mean absolute error (MAE) for ridge regression [(a), (b)] and RBF kernel ridge regression [(c), (d)] models trained on three-qubit Haar-random [(a), (c)] and VQE-ansatz [(b), (d)] datasets at different training-set sizes $N$. Solid lines show the proposed (blue) and conventional (red) ML-QEM methods (mean $\pm$ standard deviation over 10 independent seeds); dotted lines show before ML mitigation baselines.}
		\label{fig:3qubit_metrics}
\end{figure*}
We evaluate the performance of the proposed machine learning-based quantum error mitigation (ML-QEM) framework for estimating the ground-state energies and state fidelities of the three-qubit $\Lambda$-deuteron and four-qubit $\Lambda$-$\alpha$ systems. We compare the results with those obtained using conventional ML-QEM, which serves as the baseline approach. Two classical machine learning models, ridge regression and nonlinear  radial basis function (RBF) kernel ridge regression, are used to learn the corrections, and their predictive capabilities are evaluated.

\subsubsection{Predictive performance of ML models for three-qubit system}

We first evaluate how well each model predicts the ideal Pauli expectation values for held-out test states drawn from the same distribution as the training data. Fig.~\ref{fig:3qubit_metrics} shows the mean absolute error (MAE), calculated using Eq.~\eqref{eq:MAE}, between the predicted and ideal Pauli expectation values for the ridge regression and RBF kernel ridge regression models trained on the Haar-random and VQE-ansatz datasets. The MAE calculated from the noisy values (before ML correction) is also shown for reference. The root mean square error (RMSE) and coefficient of determination ($R^2$), calculated using Eqs.~\eqref{eq:RMSE} and \eqref{eq:R2}, respectively, show similar trends and lead to the same conclusions. These results are provided in the Appendix.

\paragraph{Haar-random.} For Haar-random states, the MAE of the noisy test states is $0.036$, which decreases to $0.033$ after convex projection alone. With the proposed ML-QEM framework, the MAE further decreases to $0.025$ for ridge regression and $0.027$ for RBF kernel ridge regression at $N=1600$. These values correspond to reductions of $30.7\%$ and $27.1\%$, respectively, relative to the noisy input, and $24.5\%$ and $20.6\%$ relative to projection alone. In contrast, conventional ML-QEM provides very little or no benefit.

\paragraph{VQE-ansatz.} The MAE of the raw held-out test states is $0.149$, and convex projection alone reduces it to $0.082$. With the proposed ML-QEM framework, the MAE decreases with increasing training-set size $N$ for both models. At $N=1600$, ridge achieves an MAE of $0.020$ and RBF kernel ridge regression achieves $0.017$, corresponding to reductions of $86.6\%$ and $88.6\%$, respectively,  relative to the noisy input, and $75.6\%$ and $79.3\%$ relative to projection alone. At smaller values of $N$, ridge regression performs slightly better than RBF kernel ridge regression. Conventional ML-QEM also improves the predictions compared with the noisy input for both models. However, its errors remain higher than those obtained with the proposed approach for all training-set sizes.
	
Overall, the proposed ML-QEM framework outperforms the conventional ML-QEM variant across both regression models, both training-data strategies, and all training-set sizes considered. In addition, models trained on VQE ansatz states achieve significant reduction in errors than Haar-random states.

\begin{figure*}[t]
	\centering
	\captionsetup{}
	\includegraphics[width=1\textwidth]{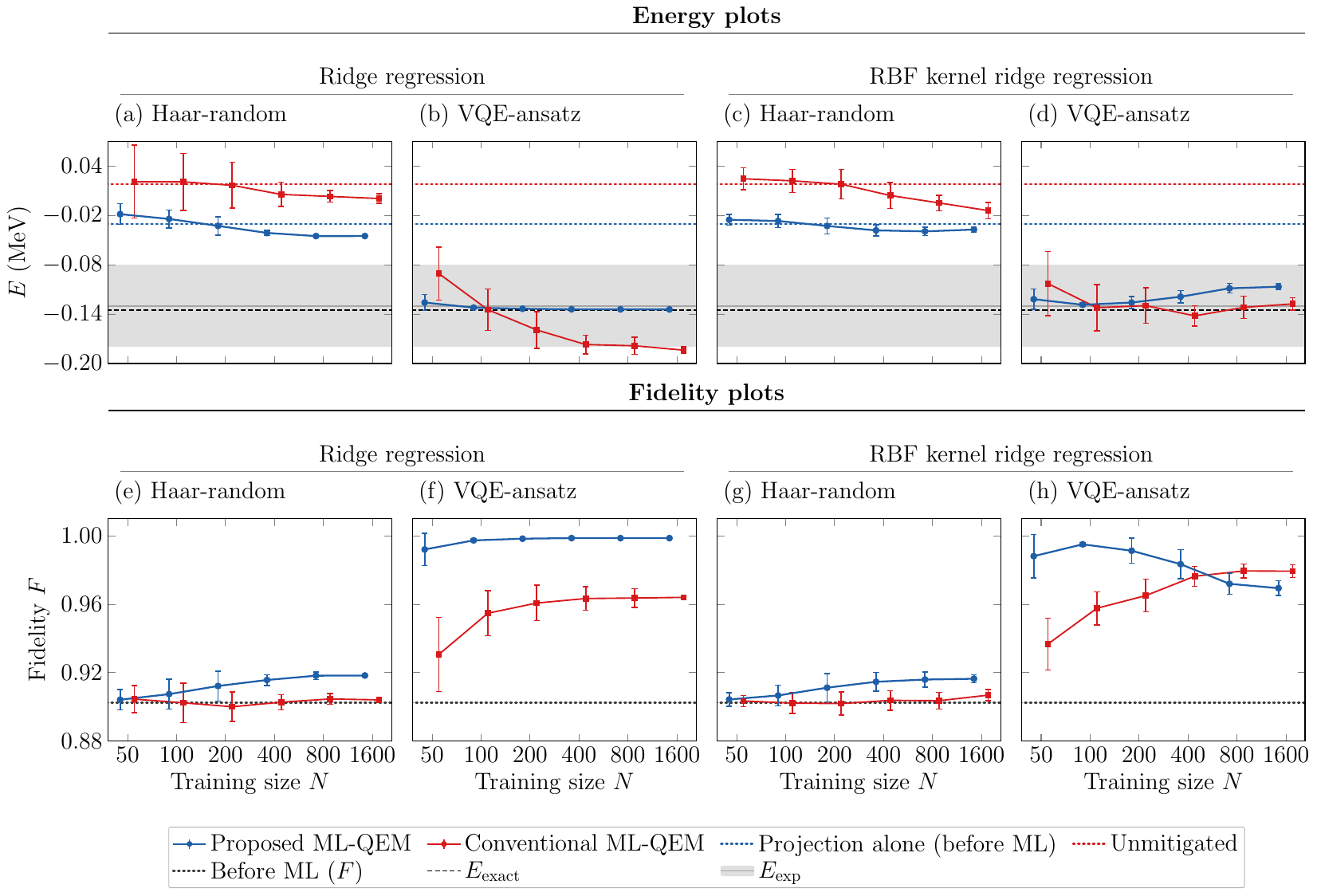}
	\caption{Ground-state energies [panels (a)--(d)] and state fidelities [panels (e)--(h)] of the three-qubit mapped $\Lambda$-deuteron system obtained with conventional (red) and proposed (blue) ML-QEM at different training-set sizes $N$. Panels (a), (b), (e), (f) show ridge regression and (c), (d), (g), (h) show RBF kernel ridge regression, each trained on Haar-random [(a), (c), (e), (g)] and VQE-ansatz [(b), (d), (f), (h)] datasets. Error bars denote the standard deviation over 10 independent seeds. Dotted lines: values before ML mitigation; dashed line: $E_{\rm exact}$; gray band: $E_{\rm exp}$. }
	\label{fig:3qubit_energy_VQE}
\end{figure*}

\begin{table*}[]
\begin{ruledtabular}
\caption{Ground-state energy estimates (in MeV) and state fidelities of the $\Lambda$-deuteron system for the Woods-Saxon potential at $N=1600$ using ridge and RBF kernel ridge regression (RBF-KRR). Results are reported as the mean $\pm$ standard deviation over 10 runs.}
\label{tab:energy_fidelity_deuteron}
\begin{tabular}{c c c c c c}
& & \multicolumn{2}{c}{Conventional ML-QEM} & \multicolumn{2}{c}{Proposed ML-QEM} \\
\cline{3-4}\cline{5-6}
Potential & Model & Haar-random & VQE-ansatz & Haar-random & VQE-ansatz \\
\hline

\multicolumn{6}{c}{\textit{Ground-State Energy}} \\
\hline
\multirow{2}{*}{Woods-Saxon}
& Ridge   & $0.000 \pm 0.006$ & $-0.184 \pm 0.004$ & $-0.045 \pm 0.001$ & $-0.134 \pm 0.000$ \\
& RBF-KRR & $-0.014 \pm 0.010$ & $-0.128 \pm 0.007$ & $-0.037 \pm 0.004$ & $-0.107 \pm 0.004$ \\
\hline

\multicolumn{6}{c}{\textit{State Fidelity}} \\
\hline
\multirow{2}{*}{Woods-Saxon}
& Ridge   & $0.904 \pm 0.001$ & $0.964 \pm 0.001$ & $0.918 \pm 0.001$ & $0.999 \pm 0.000$ \\
& RBF-KRR & $0.907 \pm 0.003$ & $0.979 \pm 0.004$ & $0.916 \pm 0.002$ & $0.969 \pm 0.004$ \\
\end{tabular}
\end{ruledtabular}
\end{table*}

\subsubsection{Three-qubit ground-state energy and fidelity analysis}

The ground-state energies and state fidelities obtained using conventional and proposed ML-QEM for the three-qubit mapped $\Lambda$-deuteron system are shown in Fig.~\ref{fig:3qubit_energy_VQE}. The unmitigated hardware energy (REM incorporated) gives an energy of $0.018$ MeV, against the ideal value of $-0.135$ MeV with a state fidelity of $0.902$. The projection alone improves the energy estimate to $E_{\rm proj}=-0.030$ MeV. Table~\ref{tab:energy_fidelity_deuteron} reports the ground-state energy estimates and state fidelities of the target state ($\Lambda$-deuteron system) at training-set size $N$=1600 for both datasets. The proposed and conventional ML-QEM calculations were repeated over 10 independent runs under the same conditions. Results are reported as the mean $\pm$ standard deviation over these runs.

\paragraph{Haar-random.} Training on Haar-random states gives smaller improvements for the target system. With the proposed ML-QEM framework, the predicted energy changes from $-0.018$ MeV at $N=50$ to $-0.045$ MeV at $N=1600$ for ridge regression and $-0.025$ MeV at $N=50$ to $-0.037$ MeV at $N=1600$ for RBF kernel ridge regression, respectively. At $N=1600$, the state fidelities reach only $0.918$ and $0.916$ for the two regression models, respectively, compared with $0.902$ before correction. With conventional ML-QEM, neither model gives an energy estimate better than the raw hardware value. At $N=1600$, the predicted energies are $0.001$ MeV for ridge regression and $-0.014$ MeV for RBF kernel ridge regression. The corresponding fidelities are $0.904$ and $0.907$, respectively.

\paragraph{VQE-ansatz.} For the proposed ML-QEM framework, ridge regression performs best. The predicted energy moves from $-0.126$ MeV at $N=50$ to $-0.134$ MeV at $N=1600$, which is close to $E_{\mathrm{ideal}}$. The corresponding state fidelity rises from $0.992$ to $0.999$. The predicted energy does not fall below the exact ground-state energy for any of the considered training-set sizes. In comparison, RBF kernel ridge regression gives its closest energy estimate of $-0.129$ MeV with state fidelity of $0.995$, at $N=100$. However, its performance decreases as the training-set size increases.

For conventional ML-QEM, the energy predicted by ridge regression moves below $E_{\mathrm{ideal}}$ as the training-set size $N$ increases. It changes from $-0.091$ MeV at $N=50$ to $-0.184$ MeV at $N=1600$, while the state fidelity remain constant near $0.96$. In comparison, RBF kernel ridge regression gives energy estimates between $-0.103$ and $-0.142$ MeV, while the fidelity increasing from $0.937$ to $0.979$.

\begin{figure*}[]
	\includegraphics[width=1\textwidth]{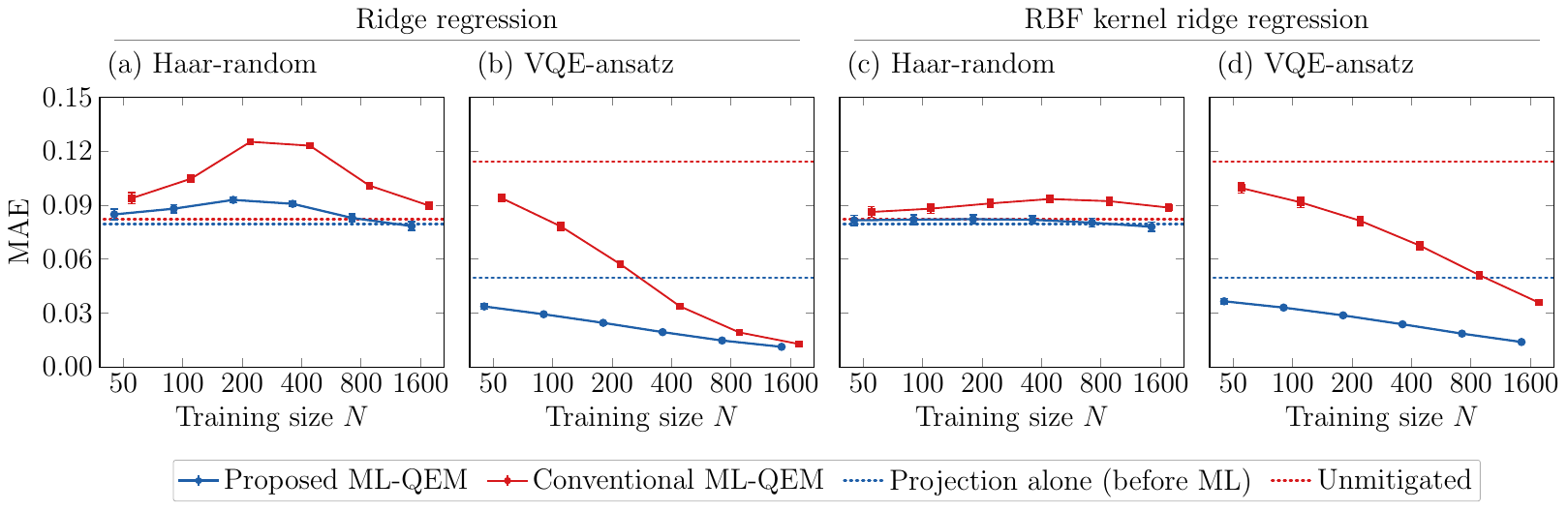}
	\caption{Test-set mean absolute error (MAE) for ridge regression [(a), (b)] and RBF kernel ridge regression [(c), (d)] models trained on four-qubit Haar-random [(a), (c)] and VQE-ansatz [(b), (d)] datasets at different training-set sizes $N$. Solid lines show the proposed (blue) and conventional (red) ML-QEM methods (mean $\pm$ standard deviation over 10 independent seeds); dotted lines show before ML mitigation baselines.}
	\label{fig:4qubit_MAE_metrics}
\end{figure*}

\subsection{Predictive performance of ML models for four-qubit system}

We examine the predictive performance of the ML models for the four-qubit system using the same procedure and metrics as for the three-qubit system. Figure~\ref{fig:4qubit_MAE_metrics} shows the test-set mean absolute error (MAE), calculated between the predicted and ideal Pauli expectation values for the ridge regression and RBF kernel ridge regression models trained on Haar-random and VQE-ansatz datasets. The same metrics evaluated on the noisy values (before ML) are reported for reference. The root mean square error (RMSE) and coefficient of determination ($R^2$) give similar trends and are provided in the Appendix.

\begin{table*}[]
\begin{ruledtabular}
\caption{Ground-state energies estimates (in MeV) and state fidelities of the $\Lambda-\alpha$ system for the Gibson~I and Bassichis-Gal \textit{c} potentials at $N=1600$ using ridge and RBF kernel ridge regression (RBF-KRR). Results are reported as the mean $\pm$ standard deviation over 10 runs.}
\label{tab:combined_energy_fidelity}
\begin{tabular}{c c c c c c}
& & \multicolumn{2}{c}{Conventional ML-QEM} & \multicolumn{2}{c}{Proposed ML-QEM} \\
\cline{3-4}\cline{5-6}
Potential & Model & Haar-random & VQE-ansatz & Haar-random & VQE-ansatz \\
\hline

\multicolumn{6}{c}{\textit{Ground-State Energy}} \\
\hline
\multirow{2}{*}{Gibson~I}
& Ridge    & $-0.636 \pm 0.366$ & $-3.544 \pm 0.068$ & $-0.495 \pm 0.098$ & $-2.663 \pm 0.026$ \\
& RBF-KRR & $-1.151 \pm 0.398$ & $-3.822 \pm 0.102$ & $-0.584 \pm 0.105$ & $-3.014 \pm 0.058$ \\
\hline
\multirow{2}{*}{Bassichis-Gal \textit{c}}
& Ridge    & $-0.346 \pm 0.329$ & $-3.095 \pm 0.070$ & $0.283 \pm 0.113$ & $-2.502 \pm 0.111$ \\
& RBF-KRR & $-1.032 \pm 0.357$ & $-3.739 \pm 0.097$ & $0.156 \pm 0.079$ & $-3.045 \pm 0.004$ \\
\hline

\multicolumn{6}{c}{\textit{State Fidelity}} \\
\hline
\multirow{2}{*}{Gibson~I}
& Ridge    & $0.865 \pm 0.005$ & $0.948 \pm 0.004$ & $0.874 \pm 0.004$ & $0.965 \pm 0.002$ \\
& RBF-KRR & $0.870 \pm 0.003$ & $0.978 \pm 0.004$ & $0.874 \pm 0.003$ & $0.995 \pm 0.004$ \\
\hline
\multirow{2}{*}{Bassichis-Gal \textit{c}}
& Ridge    & $0.842 \pm 0.004$ & $0.951 \pm 0.004$ & $0.867 \pm 0.004$ & $0.970 \pm 0.003$ \\
& RBF-KRR & $0.858 \pm 0.003$ & $0.977 \pm 0.003$ & $0.870 \pm 0.002$ & $0.998 \pm 0.000$ \\
\end{tabular}
\end{ruledtabular}
\end{table*}

\paragraph{Haar-random.} For the noisy test states, the MAE is $0.082$, which decreases to $0.079$ after convex projection alone. With the proposed ML-QEM framework, the MAE changes only slightly with training-set size $N$ and reaches $0.078$ for both ridge and RBF kernel ridge regression at $N=1600$. With conventional ML-QEM, the MAE remains higher than that of the noisy input at every training-set size. At $N=1600$, it is $0.090$ for ridge regression and $0.089$ for RBF kernel ridge regression.

\paragraph{VQE-ansatz.} The MAE of the noisy test data is $0.114$, and convex projection alone reduces it to $0.050$. With the proposed ML-QEM framework, both models show improvement with increase in training set size $N$. At $N=1600$, ridge regression gives an MAE of $0.011$, while RBF kernel ridge regression gives $0.014$. These values correspond to the reduction of $90.3\%$ and $87.8\%$, respectively, with respect to the noisy input, and $78\%$ and $72.0\%$ with respect to the projection alone. With conventional ML-QEM, both models also improve upon the noisy input, but they remain less accurate than the proposed approach at every $N$. At $N=1600$, the MAE is $0.013$ for ridge regression and $0.036$ for RBF kernel ridge regression.

Overall, the proposed ML-QEM framework gives lower prediction errors than conventional ML-QEM for both regression models, both training datasets, and all training-set sizes considered. Furthermore, models trained on VQE-ansatz states achieve substantially larger error reductions than those trained on Haar-random states.

\begin{table*}[]
\caption{Ground-state energy estimates (in MeV) for the three-qubit $\Lambda$-deuteron system with the Woods-Saxon potential. The ideal ground-state energy is compared with hardware results obtained without mitigation (raw), with readout error mitigation (REM), and with zero-noise extrapolation combined with REM (ZNE+REM). The ML-QEM results are obtained using ridge regression and RBF kernel ridge regression (RBF-KRR), trained on VQE-ansatz data with $N=1600$, for both conventional and proposed approaches. ML-QEM results are reported as the mean $\pm$ standard deviation over 10 seeds.}
\label{tab:summary_energy}
\begin{ruledtabular}
\begin{tabular}{lcccccccc}
\multirow{2}{*}{Potential} & \multirow{2}{*}{Ideal} & \multirow{2}{*}{Raw} & \multirow{2}{*}{REM} & \multirow{2}{*}{ZNE+REM} & \multicolumn{2}{c}{Conventional ML-QEM} & \multicolumn{2}{c}{Proposed ML-QEM} \\
\cline{6-7}\cline{8-9}
& & & & & Ridge & RBF-KRR & Ridge & RBF-KRR \\
\hline
Woods-Saxon
& $-0.135$
& $0.618$
& $-0.001$ & $-0.046$ & $-0.184$ & $-0.128$ & $-0.134$ & $-0.107$ \\
& & 
& $\pm 0.024$ & $\pm 0.022$ & $\pm 0.004$ & $\pm 0.007$ & $\pm 0.000$ & $\pm 0.004$ \\
\end{tabular}
\end{ruledtabular}
\end{table*}

\begin{table*}[]
\caption{Ground-state energy estimates (in MeV) for the four-qubit $\Lambda$-$\alpha$ system using the Gibson~I and Bassichis-Gal \textit{c} potentials. The ideal ground-state energies are compared with hardware results obtained without mitigation (raw), with readout error mitigation (REM), and with zero-noise extrapolation combined with REM (ZNE+REM). The ML-QEM results are obtained using ridge regression and RBF kernel ridge regression (RBF-KRR), trained on VQE-ansatz data with $N=1600$, for both conventional and proposed approaches. ML-QEM results are reported as the mean $\pm$ standard deviation over 10 seeds.}
\label{tab:4qubit_summary_energy}
\begin{ruledtabular}
\begin{tabular}{lcccccccc}
\multirow{2}{*}{Potential} & \multirow{2}{*}{Ideal} & \multirow{2}{*}{Raw} & \multirow{2}{*}{REM} & \multirow{2}{*}{ZNE+REM} & \multicolumn{2}{c}{Conventional ML-QEM} & \multicolumn{2}{c}{Proposed ML-QEM} \\
\cline{6-7}\cline{8-9}
& & & & & Ridge & RBF-KRR & Ridge & RBF-KRR \\
\hline
\multirow{2}{*}{Gibson~I}
& \multirow{2}{*}{$-3.118$}
& \multirow{2}{*}{$6.851$}
& $-0.568$ & $-1.125$ & $-3.544$ & $-3.822$ & $-2.663$ & $-3.014$ \\
& & & $\pm 0.343$ & $\pm 0.412$ & $\pm 0.068$ & $\pm 0.102$ & $\pm 0.026$ & $\pm 0.058$ \\
\hline
\multirow{2}{*}{Bassichis-Gal \textit{c}}
& \multirow{2}{*}{$-3.115$}
& \multirow{2}{*}{$6.569$}
& $0.391$ & $-0.672$ & $-3.095$ & $-3.739$ & $-2.502$ & $-3.045$ \\
& & & $\pm 0.570$ & $\pm 0.676$ & $\pm 0.070$ & $\pm 0.097$ & $\pm 0.111$ & $\pm 0.004$ \\
\end{tabular}
\end{ruledtabular}
\end{table*}

\subsection{Four-qubit ground-state energy and fidelity analysis}

\begin{figure*}
	\centering
	\captionsetup{}
	\includegraphics[width=1\textwidth]{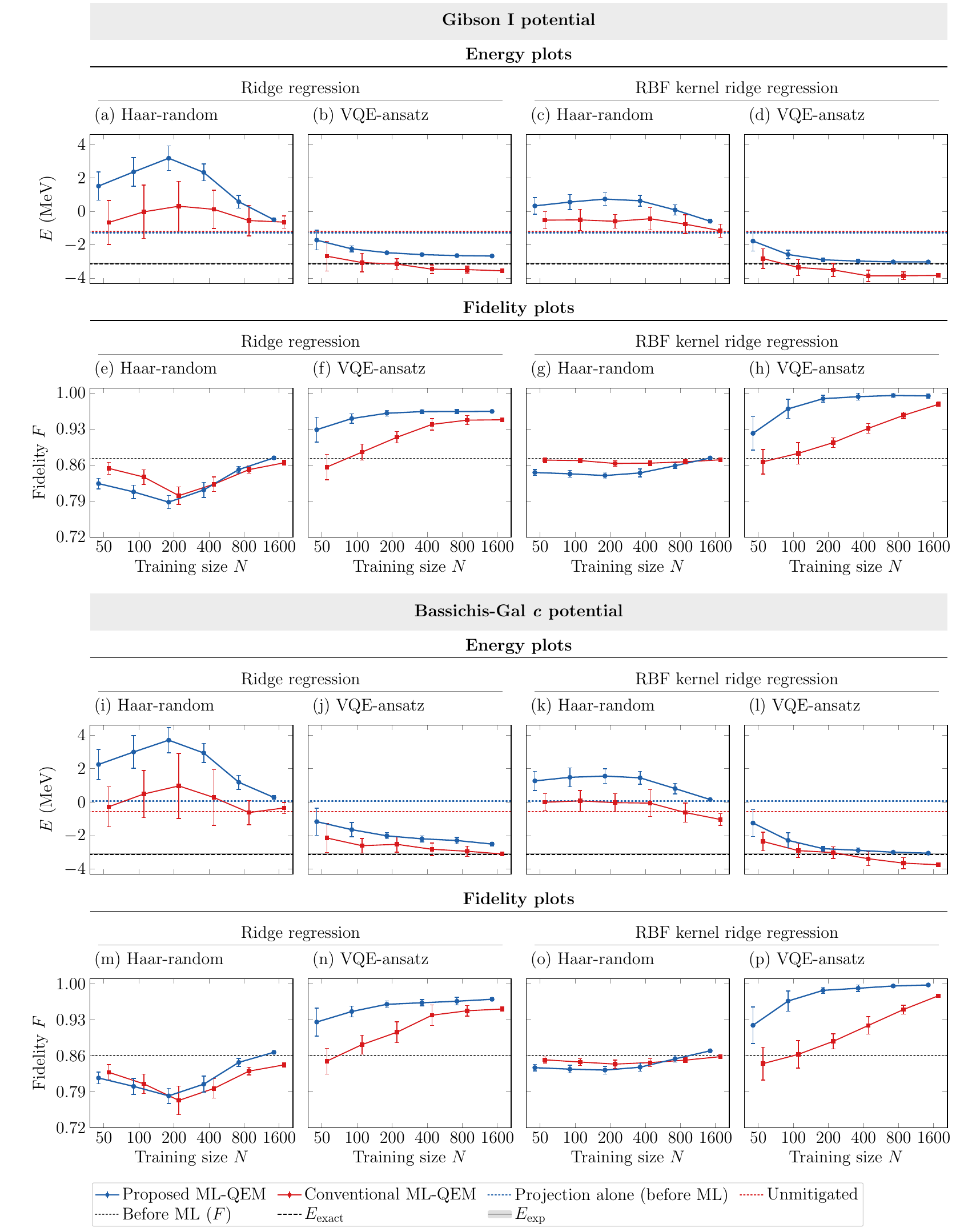}
	\caption{Ground-state energies [(a)--(d), (i)--(l)] and state fidelities [(e)--(h), (m)--(p)] for the four-qubit $\Lambda$-$\alpha$ system, comparing  conventional (red) and proposed (blue) ML-QEM, for the Gibson~I and Bassichis-Gal \textit{c} potentials. In each row, the left two panels show ridge regression and the right two RBF kernel ridge regression, each trained on Haar-random and VQE-ansatz datasets. Error bars denote the standard deviation over 10 independent runs. Dotted lines: values before ML mitigation; dashed line: $E_{\rm exact}$; gray band: $E_{\rm exp}$.}
	\label{fig:4qubit_energy1}
\end{figure*}

The ground-state energies and state fidelities obtained using the proposed ML-QEM framework for the four-qubit mapped $\Lambda$-$\alpha$ system with the Gibson~I and Bassichis-Gal \textit{c} potentials are shown in Fig.~\ref{fig:4qubit_energy1}. Compared with the three-qubit system, the four-qubit system has a larger Hilbert space and the increased number of Pauli observables. The ideal ground-state energies are $E_{\rm ideal}=-3.118$~MeV and $-3.115$~MeV for the Gibson~I and Bassichis-Gal \textit{c} potentials, respectively.  Before ML mitigation, the noisy hardware energies (REM incorpotated) are $-1.183$ MeV and $-0.565$ MeV, respectively. Convex projection alone gives energy estimates of $-1.274$ MeV and $0.060$ MeV, while the corresponding state fidelities of the noisy reconstructed states are $0.873$ and $0.861$.

\paragraph{Haar-random.} Training on Haar-random states does not give accurate ground-state energy estimates. For both potentials and both regression models, the predicted energies remain far above $E_{\rm ideal}$. At $N=1600$, the proposed ML-QEM framework gives $-0.495$ MeV for ridge regression and $-0.584$ MeV for RBF kernel ridge regression for Gibson~I potential. For the Bassichis-Gal \textit{c} potential, the corresponding energies are $0.283$ MeV and $0.156$ MeV. All these values are higher than the corresponding unmitigated energies.

For conventional ML-QEM, the corresponding energies at $N=1600$ are $-0.636$ MeV and $-1.151$ MeV for the Gibson~I potential, and $-0.346$ MeV and $-1.032$ MeV for the Bassichis-Gal~\textit{c} potential, using ridge and RBF kernel ridge regression, respectively. The state fidelities show similarly limited improvements for both proposed and conventional ML-QEM, remaining close to their initial values even at the largest training-set size.

\paragraph{VQE-ansatz.} With the proposed ML-QEM framework, both regression models give energy estimates close to the exact ground-state energies. At $N=1600$, RBF kernel ridge regression gives $-3.014$ MeV for Gibson~I and $-3.045$ MeV for the Bassichis-Gal \textit{c} potential. In comparison, ridge regression yields $-2.663$ MeV and $-2.502$ MeV, respectively. The state fidelity increases with $N$ and reaches $0.995$ and $0.998$ for RBF kernel ridge regression, and $0.965$ and $0.970$ for ridge regression, for the Gibson~I and Bassichis-Gal~\textit{c} potentials, respectively.

In contrast, conventional ML-QEM with RBF kernel ridge regression produces energies below the exact ground-state values, reaching $-3.822$ MeV and $-3.739$ MeV for the Gibson~I and Bassichis-Gal~\textit{c} potentials, respectively. Ridge regression also gives an energy below the exact value for the Gibson~I potential ($-3.544$ MeV). However, for the Bassichis-Gal~\textit{c} potential, it yields $-3.095$ MeV, which is closer to the exact value than the result obtained using the proposed ridge regression approach. The state fidelity obtained with conventional ML-QEM increases more slowly than that obtained with the proposed approach. At $N=1600$, the conventional state fidelities reach $0.948$ and $0.951$ for ridge regression, and $0.978$ and $0.977$ for RBF kernel ridge regression, for the Gibson~I and Bassichis-Gal \textit{c} potentials, respectively. These values remain below those obtained with the proposed ML-QEM.

Overall, these results indicate that Haar-random training data do not provide effective corrections for the target ground states, particularly when compared with the VQE-ansatz training data.

\subsection{Overall comparison and discussion}

The results show that the performance of ML-QEM depends on both the choice of training dataset and the regression model used to learn the correction. For both system sizes, models trained on VQE-ansatz states give lower Pauli-prediction errors than those trained on Haar-random states. The same trend is observed in the ground-state calculations. VQE-ansatz training data give more accurate energy estimates and higher state fidelities, whereas Haar-random training generally does not recover the target ground states.

Tables~\ref{tab:summary_energy} and~\ref{tab:4qubit_summary_energy} compare the ground-state energy estimates obtained using different error-mitigation approaches. The results show that conventional error-mitigation techniques alone are insufficient to recover the exact ground-state energies. Although readout error mitigation (REM) and zero noise extrapolation (ZNE) combined with REM reduce the errors in the hardware estimates, substantial deviations from the ideal values remain. In comparison, all the ML-QEM approaches considered give energy estimates closer to the exact values than the conventional error-mitigation baselines.

Using the proposed ML-QEM framework with VQE-ansatz training data, we obtain the most accurate ground-state energy for both the three- and four-qubit $\Lambda$-deuteron and $\Lambda$-$\alpha$  system. For the three-qubit $\Lambda$-deuteron system, ridge regression gives the best result at $N=1600$. The predicted energy is $-0.134$ MeV, which is in close agreement with the exact value of $-0.135$ MeV, with a corresponding state fidelity of 0.999. For the four-qubit $\Lambda$-$\alpha$  system, the RBF kernel ridge regression model gives the best results for both nuclear potentials. It predicts ground-state energies of $-3.014$ MeV and $-3.045$ MeV for the Gibson I and Bassichis-Gal \textit{c} potentials, respectively, with corresponding state fidelities of $0.995$ and $0.998$. These results show that the choice of regression model depends on the system being studied, with ridge regression performing best for the $\Lambda$-deuteron and RBF kernel ridge regression for the $\Lambda$-$\alpha$ system.

The comparison with conventional ML-QEM further shows that increasing the training-set size does not necessarily lead to better predictions. In particular, conventional ridge for the three-qubit system and conventional RBF kernel ridge regression for the four-qubit system produce energy estimates below the exact ground-state values.  Furthermore, although convex projection alone improves the physical validity of the reconstructed states, but does not consistently recover the ground-state energy. Overall, the results show the importance of using physically constrained quantum-state reconstruction, training data that represent the target states, and a suitable regression model.

\section{\label{5}Conclusions and Outlook}

In this work, we investigated machine learning-based quantum error mitigation (ML-QEM) for ground-state energy estimation and quantum-state reconstruction in the three-qubit $\Lambda$-deuteron and four-qubit $\Lambda$-$\alpha$ nuclear systems using an IBM Quantum processor. The proposed approach learns the residual correction between ideal and noisy Pauli expectation values and projects the corrected result onto the set of physical quantum states. We compared its performance with conventional ML-QEM and the baseline error-mitigation techniques, readout error mitigation (REM) and zero-noise extrapolation (ZNE). We also examined the effects of the training-state distribution, training-set size, and regression model on the mitigation performance.

Our results show that the proposed ML-QEM framework improves both the ground-state energy estimates and state fidelities, with the largest improvements obtained using VQE-ansatz training data. However, the performance depends on both the quantum system and the choice of regression model. For the three-qubit $\Lambda$-deuteron system, ridge regression gives stable energy estimates close to the ideal ground-state energy when trained on VQE-ansatz states. In contrast, for the four-qubit $\Lambda$-$\alpha$ system, nonlinear RBF kernel ridge regression gives better results than ridge regression for both nuclear potentials, in terms of ground-state energy and state fidelity. The results also show that the choice of training data plays an important role in ML-QEM performance. In particular, the improvements obtained with Haar-random training data are smaller than those obtained with VQE-ansatz data. These findings highlight the importance of selecting an appropriate training-state distribution and regression model, along with enforcing physical constraints on the reconstructed quantum states.

In future work, we will focus on improving the scalability of the framework and extend it to larger, more realistic many-body nuclear Hamiltonians. We will also explore other machine learning models, such as deep neural networks and graph neural networks, for predicting the ground-state properties and other observables. These studies should clarify how machine learning can support practical quantum simulations of complex nuclear systems on current noisy quantum hardware.

\begin{acknowledgments}
S. C. acknowledges IUAC (AUCID: 77316), New Delhi, for financial support. H. S. acknowledges financial support from NQM (DST/QTC/NQM/QA/2026/01(G)) for this work. 
\end{acknowledgments}

\section*{Data Availability}
Data supporting the findings of this study has been deposited in the Zenodo repository \cite{yourdataset2026} and is available at \url{https://zenodo.org/records/23969830}.

\appendix

\section*{Appendix: Additional predictive-performance metrics}
\label{6}

\begin{figure*}[]
    \includegraphics[width=1\textwidth]{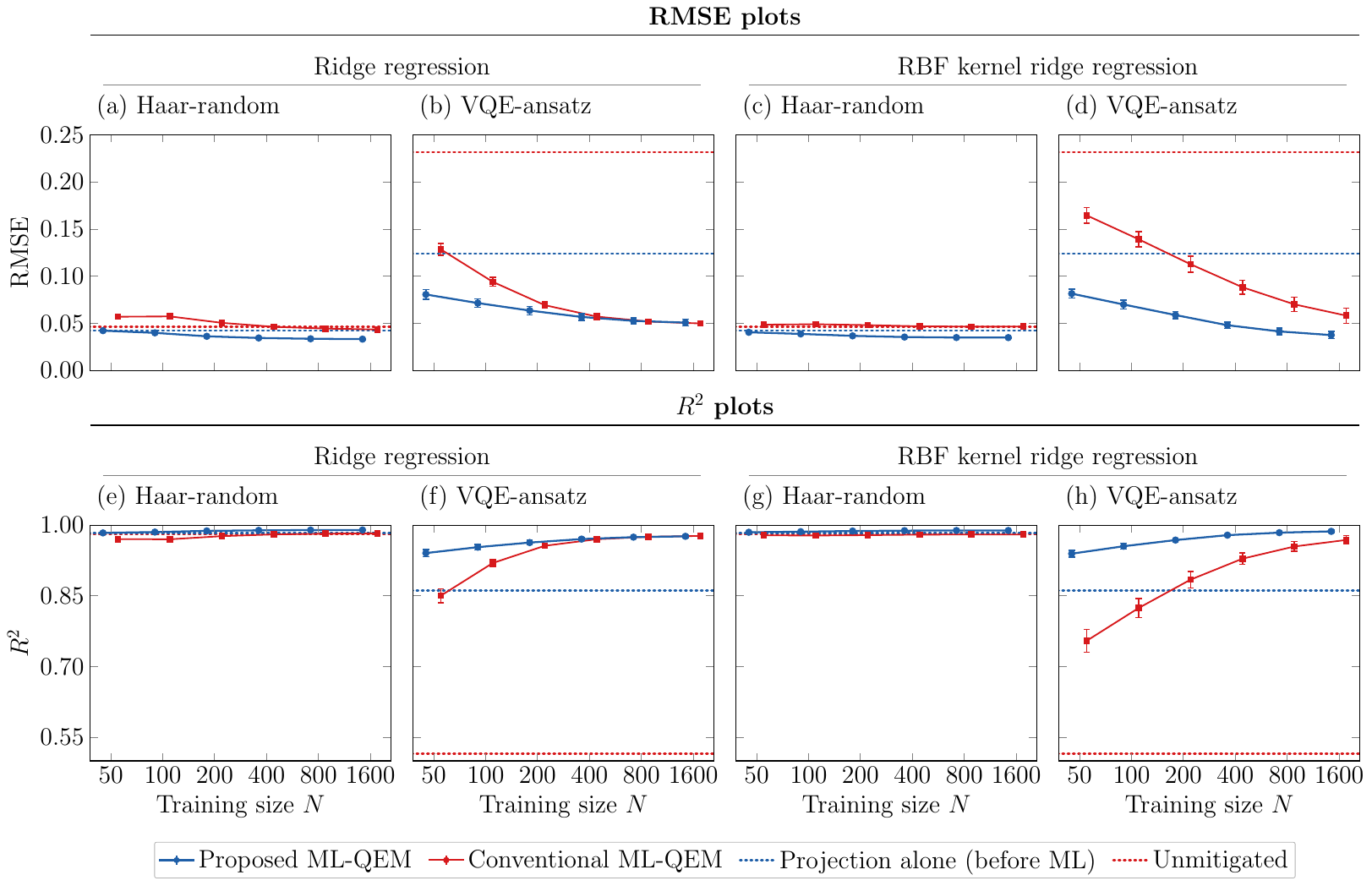}
    \caption{Test-set root mean square error (RMSE) [panels (a)--(d)] and coefficient of determination ($R^2$) [panels (e)--(h)] for ridge regression [(a), (b), (e), (f)] and RBF kernel ridge regression [(c), (d), (g), (h)] models trained on three-qubit Haar-random [(a), (c), (e), (g)] and VQE-ansatz [(b), (d), (f), (h)] datasets at different training-set sizes $N$. Solid lines show the proposed (blue) and conventional (red) ML-QEM approaches (mean $\pm$ standard deviation over 10 independent seeds); dotted lines show the values before ML mitigation.}
    \label{fig:3qubit_RMSE_metrics}
\end{figure*}

\begin{figure*}[]
    \includegraphics[width=1\textwidth]{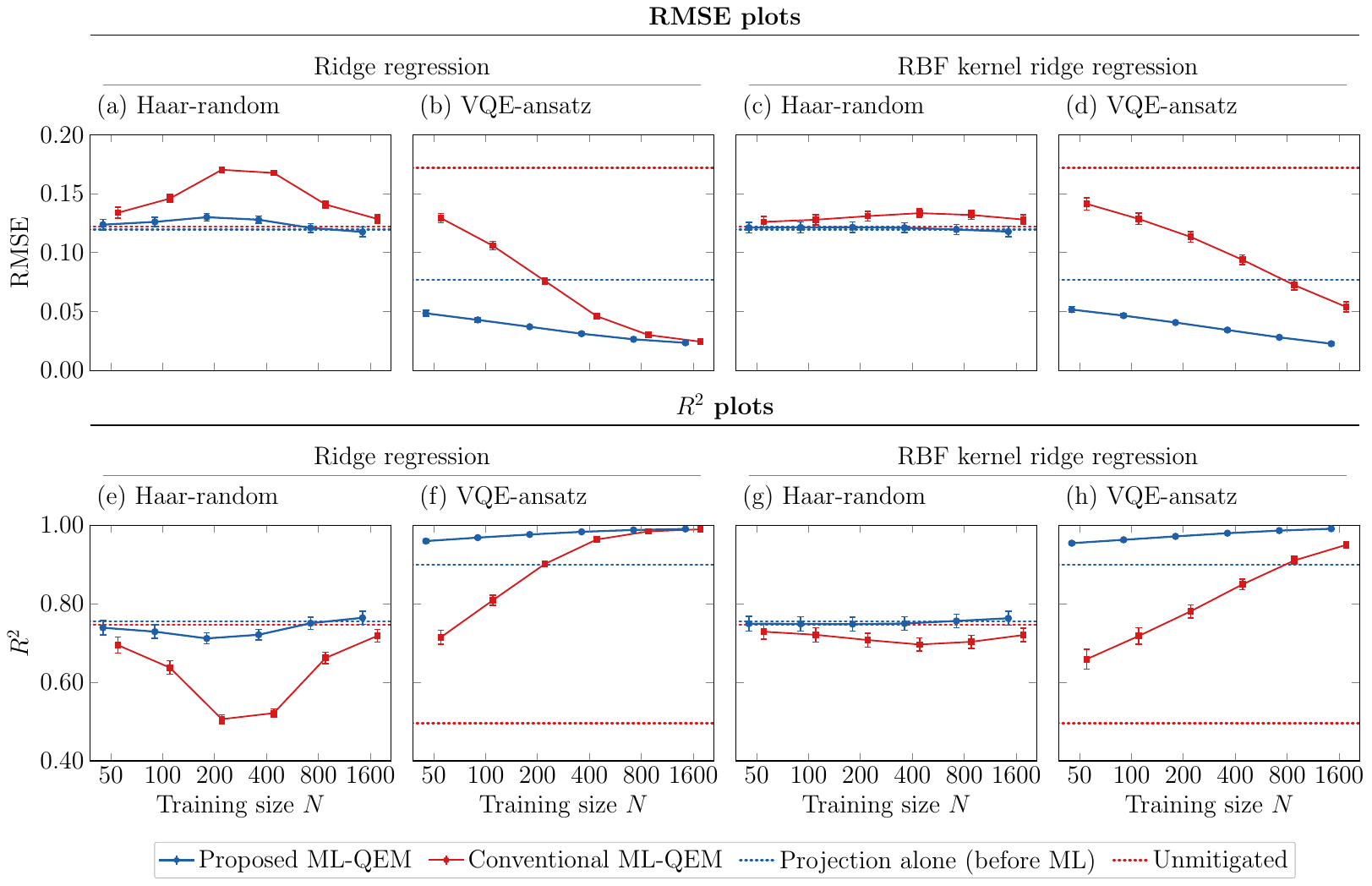}
    \caption{Test-set root mean square error (RMSE) [panels (a)--(d)] and coefficient of determination ($R^2$) [panels (e)--(h)] for ridge regression [(a), (b), (e), (f)] and RBF kernel ridge regression [(c), (d), (g), (h)] models trained on four-qubit Haar-random [(a), (c), (e), (g)] and VQE-ansatz [(b), (d), (f), (h)] datasets at different training-set sizes $N$. Solid lines show the proposed (blue) and conventional (red) ML-QEM approaches (mean $\pm$ standard deviation over 10 independent seeds); dotted lines show the values before ML mitigation.}
    \label{fig:4qubit_RMSE_metrics}
\end{figure*}

\subsection{Three-qubit system}

Fig.~\ref{fig:3qubit_RMSE_metrics} presents the root mean square error (RMSE) and coefficient of determination ($R^2$) on the held-out test states, calculated between the predicted and ideal Pauli expectation values for the regression models trained on the VQE and Haar-random datasets, complementing the MAE results discussed in the main text. The same metrics evaluated on the noisy values (before ML) are reported for reference. These additional metrics provide a further assessment of the predictive accuracy of the ridge and RBF kernel ridge regression models for the Haar-random and VQE-ansatz datasets as a function of the training set size $N$.

\paragraph{Haar-random.} The noisy input has an RMSE of $0.046$ ($R^2=0.981$), and convex projection alone lowers it to $0.042$ ($R^2=0.984$). The proposed ML-QEM reduces the error further. At $N=1600$, ridge regression reaches an RMSE of $0.033$ ($R^2=0.990$) and RBF kernel ridge regression reaches $0.035$ ($R^2=0.989$). Conventional ML-QEM, by comparison, improves only slightly on the unmitigated input.

\paragraph{VQE ansatz.} The noisy input has an RMSE of $0.232$ ($R^2=0.515$), which is reduced to $0.124$ ($R^2=0.861$) by convex projection alone. With the proposed ML-QEM framework, the RMSE decreases steadily with increasing $N$, reaching $0.051$ for ridge and $0.038$ for RBF kernel ridge regression at $N=1600$, with corresponding $R^2$ values of $0.977$ and $0.987$, respectively. The RBF kernel ridge regression model with the proposed ML-QEM consistently outperforms its conventional counterpart across all training-set sizes.

\subsection{Four-qubit system}

Fig.~\ref{fig:4qubit_RMSE_metrics} shows the corresponding RMSE and $R^2$ results for the four-qubit system, allowing the predictive performance of the two ML-QEM approaches to be compared with that observed for the three-qubit system.

\paragraph{Haar-random.} For Haar-random states, the noisy input has an RMSE of $0.122$ ($R^2=0.746$), while convex projection alone provides only a marginal improvement, yielding an RMSE of $0.120$ ($R^2=0.756$). The proposed ML-QEM also provides limited improvement over these baselines. Conventional ML-QEM performs even less favorably, yielding higher RMSE and lower $R^2$ values than the noisy input across all training-set sizes.

\paragraph{VQE ansatz.} The noisy input has an RMSE of $0.172$ ($R^2=0.496$), which is reduced to $0.077$ ($R^2=0.899$) by convex projection alone. With the proposed ML-QEM framework, both models exhibit steady improvements with increasing training-set size $N$. At $N=1600$, ridge and RBF kernel ridge regression both achieve an RMSE of $0.023$, with corresponding $R^2$ values of $0.991$ for both models. The proposed ML-QEM consistently outperforms the conventional approach for both models across all training-set sizes $N$.

\end{document}